\documentclass[12pt]{iopart}
\usepackage{graphicx}

\newcommand{\pp}{p^{\prime}}
\newcommand{\kk}{k^{\prime}}
\newcommand{\xp}{x^{\prime}}

\newcommand{\plabel}[1]
{\newcounter{#1}
\setcounter{#1}{\value{eqnval}}
}
\newcommand{\pref}[1]{(\arabic{#1})}

\begin{document}

\title{Electronic transport in inhomogeneous quantum wires}

\smallskip

\author{J\'er\^ome Rech and K. A. Matveev} 
\address{Materials Science
  Division, Argonne National Laboratory, Argonne, Illinois 60439, USA}
\eads{\mailto{rech@anl.gov}}
\submitto{\JPCM}
\begin{abstract}
  
We study the transport properties of a long non-uniform quantum wire
where the electron-electron interactions and the density vary smoothly
at large length scales. We show that these inhomogeneities lead to
a finite resistivity of the wire, due to a weak
violation of momentum conservation in the collisions between
electrons. Estimating the rate of change of momentum associated with
non-momentum-conserving scattering processes, we derive the expression
for the resistivity of the wire in the regime of weakly interacting
electrons and find a contribution linear in temperature for a broad
range of temperatures below the Fermi energy. By estimating the energy
dissipated throughout the wire by low-energy excitations, we then
develop a different method for deriving the resistivity of the wire,
which can be combined with the bosonization formalism. This allows us
to compare our results with previous works relying on an extension of
the Tomonaga-Luttinger model to inhomogeneous systems.

\end{abstract}
\maketitle

\section{Introduction}\label{intro}

Recent experiments on quantum wires and carbon nanotubes
\cite{vanwees,wharam,thomas1,thomas2,thomas3,kristensen,reilly,
cronenwett,depicciotto,rokhinson,crook,bockrath,tunneling,yao,spin-charge}
have stimulated a lot of interest in the transport properties of
one-dimensional conductors. From a theoretical point of view,
interacting electrons in one dimension form the so-called Luttinger
liquid \cite{Haldane,LL-th}, whose properties qualitatively differ
from the conventional Fermi liquid state. Recent progress in
fabrication techniques has made possible the experimental observation
of various characteristic signatures of the Luttinger liquid, such as
the power-law behavior of the tunneling density of states
\cite{bockrath,tunneling,yao}, or the existence of separate spin and
charge excitations \cite{spin-charge}. It is also expected, within the
Luttinger liquid theory, that the dc conductance of a quantum wire
connected to Fermi liquid leads is given by the quantum of conductance
$G_0 = 2 e^2/h$ \cite{maslov,safi,pono}. This quantization of the
conductance has been reported in various experimental setups since its
first observation in a quantum point contact \cite{vanwees,wharam}.

However, in a number of recent experiments \cite{thomas1,thomas2,
thomas3,kristensen,reilly,cronenwett,depicciotto,rokhinson,crook},
significant deviations from perfect quantization have been observed in
the regime of low electron density. These deviations take the form of
a shoulder-like structure below the first plateau of
conductance. Although weak at the lowest temperatures available, this
feature becomes more significant as the temperature increases, turning
into a quasi-plateau at about $0.7\times(2 e^2/h)$. This so-called
``0.7 structure'', which is not expected in the Luttinger liquid
theory, generated much theoretical interest, though there is at
present no generally accepted microscopic theory. Most commonly, the
experimental results are interpreted as originating from a
spin-dependent mechanism. Such scenarios rely on a spontaneous spin
polarization of the wire \cite{thomas1,wang,spivak}, or on the
existence of a local spin-degenerate quasi-bound state whose screening
would lead to Kondo-like effects \cite{kondo,rekondo}.  Other
proposals considered various scattering mechanisms involving plasmons
\cite{bruus}, spin waves \cite{tokura} or phonons
\cite{e-phonon}. Several authors have also suggested that
electron-electron interactions may affect the transport properties in
quantum wire devices in a way that would be consistent with the ``0.7
structure'' \cite{meidan,olav,sushkov,lunde}.

In this context, a number of recent theory papers studied the
electronic transport in a quantum wire modeled as a one-dimensional
system in which the interactions are limited to a small region
between two non-interacting leads. They concluded that the
backscattering of either single electrons or pairs were the only
mechanisms to significantly affect the transport properties of the
system \cite{sushkov,lunde}, but only if the size of the interacting
region is comparable to the Fermi wavelength of the electrons in the
wire. If, on the other hand, the interaction strength varies smoothly
over a much larger distance, such backscattering processes only lead
to exponentially small contributions which can be neglected. Using the
model of a non-uniform Luttinger liquid with position-dependent
parameters, it was found that no correction to the quantized
conductance of the wire arises in this regime
\cite{maslov,safi,pono}.

In this paper we show that even when the backscattering processes can
be ignored, the non-uniformity of the interaction potential
throughout the wire leads to a finite resistivity at non-zero
temperatures. Indeed, the inhomogeneity of the interaction potential
breaks the translational invariance of the system, allowing for
two-particle scattering processes that conserve energy but not
momentum. In \sref{qualitative}, we qualitatively show how some of
these processes give rise to a finite resistivity and perform the
corresponding calculation in \sref{stationary}. In \sref{moving}, we
present an alternative derivation of the resistivity in the language of
the inhomogeneous Luttinger liquid model, allowing us to compare our
results with previous works relying on this formalism
\cite{maslov,safi,pono}. Finally, in \sref{discussion} we discuss the
relation of our results to the experiments probing the transport
properties of inhomogeneous quantum wires. A brief summary of some of
our results was reported in \cite{letter}.

\section{Qualitative picture}\label{qualitative}

Let us consider an infinite one-dimensional system of weakly
interacting electrons with a quadratic dispersion
$\epsilon_p = p^2/2m$. To develop a qualitative picture of the physics
involved, we restrict ourselves to the simple model of spinless
electrons, with a uniform density $n$ throughout the device. (We will
tackle more realistic systems in the next section.) The inhomogeneity
of the system comes from the electron-electron interaction whose
strength varies smoothly along the wire.

When one enforces a dc current $I$ to flow through the device, the
electrons start moving and acquire a drift velocity $v_d$ proportional
to this applied current: $v_d=I/ne$. In the reference frame moving
with velocity $v_d$ along the wire, the electronic subsystem is in an
equilibrium state characterized by a Fermi energy $\epsilon_F$ and a
temperature $T$. This was recently pointed out \cite{drag} in the
context of Coulomb drag between two parallel wires.

As we are interested in the low-energy properties of the system, we
focus on temperatures $T \ll \epsilon_F$, so that the only relevant
excitations are close to the Fermi level. As a result, one can isolate
two well-defined branches corresponding to two species of fermions:
the right- and left-moving electrons. Within each branch, the velocity
of the electrons can be approximated by a constant and is given by
$+v_F$ and $-v_F$ respectively for right- and left-movers. Upon
changing from the moving to the stationary frame of reference, the
electron velocities are modified in order to account for the drift
velocity, and change from $\pm v_F$ to $\pm v_F + v_d$. The
consequences for the electron fluid as described in the stationary
frame of reference are two-fold. First, we need to introduce different
Fermi energies for right- and left-moving electrons, $\epsilon_F \to
\epsilon_F^{R,L} = (1/2) m (v_F \pm v_d)^2$. Second, since the density
of states at the Fermi level is inversely proportional to velocity, we
now have different densities of states for the two subsystems, $\nu
\propto 1/v_F \to \nu_{R,L} \propto 1/(v_F \pm v_d)$.

The latter result implies that the energy spacing between states is
not only modified as we change the frame of reference, but also
differs between the right and left branches in the stationary
frame. Compared to the moving frame, the energy levels are stretched
near the right Fermi point. This results in a somewhat broader
distribution function, which can be interpreted as a slightly higher
effective temperature $T_R$ for the right-moving electrons (see
\fref{fig2}). Similarly, near the left Fermi point, the energy levels
are squeezed compared to the moving frame, resulting in a narrower
distribution function, corresponding to a lower effective temperature
$T_L$ for the left-moving electrons. These effective temperatures
follow the change in the density of states and are given by:
\begin{equation}
T_{R,L} = T \left(1 \pm \frac{v_d}{v_F} \right) = T \left( 1 \pm \frac{I}{e n v_F} \right) .  \label{trl}
\end{equation}
The nature of these effective temperatures can be understood formally,
by noticing that in the stationary frame, the system is no longer in
thermal equilibrium because of the finite electric current. It follows
that, quite generally, the occupation probability of a given state is
no longer given by the standard Fermi-Dirac distribution. However, the
introduction of the effective temperatures \eref{trl} for right- and
left-movers enables one to write their occupation probabilities as Fermi
functions of energy.

\begin{figure}
\begin{center}
\includegraphics[width=12cm]{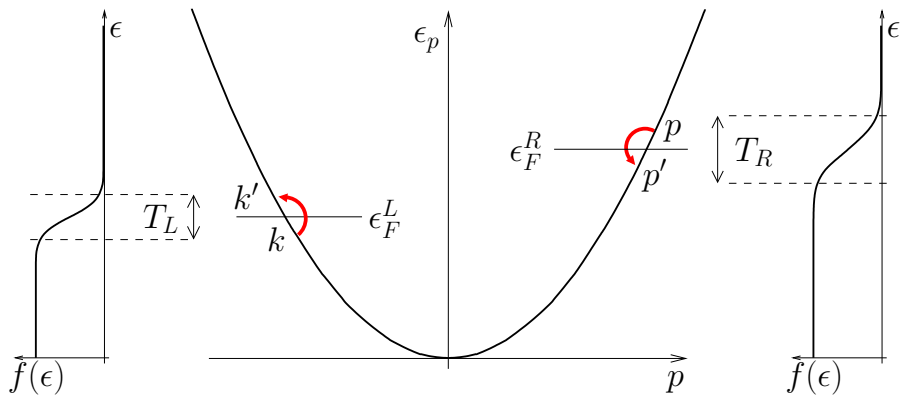}
\end{center}
\caption{
Electronic spectrum in the stationary frame, with the Fermi energies
and effective temperatures for the right- and left-moving
branches. The corresponding distribution functions near the right and
left Fermi points are displayed as functions of energy. An example of
non-momentum-conserving scattering process is provided.}
\label{fig1}
\end{figure}

Because right- and left-movers have different temperatures, it is
natural to expect that electron-electron interactions will give rise
to thermalization between the two branches. In a uniform system,
two-particle scattering processes cannot lead to thermalization as the
conservation of both energy and momentum only allows processes which
either exchange the momenta of the two incoming electrons or leave
them unchanged \cite{3electron}. On the other hand, in the case of
inhomogeneous wires, the strength of the interaction potential is
non-uniform so that the system is no longer translationally invariant,
and two-particle scattering processes which conserve energy but not
momentum are allowed.

A typical example of such electron-electron scattering processes is
shown in \fref{fig1}. It describes the scattering of two electrons
from an initial state with momenta $p$ and $k$ to a final state with
momenta $\pp$ and $\kk$, and violates the momentum conservation:
$\pp+\kk-p-k= P < 0$. Though the loss of momentum associated with this
scattering process may affect the transport properties of the system,
one could argue that it is compensated by an equal gain of momentum
corresponding to the inverse process $(\pp,\kk) \to (p,k)$.  This is
however not the case here because of the temperature difference
between the two branches: the processes involving a transfer of energy
from the ``warmer'' right-moving branch to the ``colder'' left-moving
one statistically occur more often than the corresponding inverse
processes. As a result, the electronic system loses more momentum than
it gains.

This overall loss of momentum can be viewed as resulting from a
damping force, associated with the electron-electron collisions, and
proportional to the temperature difference between the right- and
left-moving branches. In order for a constant current to flow through
the wire, this damping force has to be compensated by a driving
force. The latter originates from a local electric field which appears
as a response of the system to the external current. Using the force
balance, and keeping in mind that the temperature difference $T_R-T_L
\propto I$, this local electric field is proportional to
the applied current bias. This implies a finite resistivity of the
wire.

\section{Weakly interacting electrons in the stationary frame}
\label{stationary}

The above arguments provide a physical picture of how
inhomogeneities lead to a finite resistivity. We now proceed with the
calculation of the resistivity.

\subsection{Model} \label{model}

Our starting point is a one-dimensional system of weakly interacting
electrons with spins. In order to account for a non-uniform electron
density $n(x)$, we introduce a one-particle potential $U(x)$
originating from the surrounding gates and impurities in the
substrate. Moreover, the interaction between electrons is
inhomogeneous, and described by a smoothly varying potential $V(r,R)$,
given in the center-of-mass coordinates. The Hamiltonian for this
system takes the form
\numparts
\plabel{hf}
\begin{eqnarray}
H   = & H_0 + H_{\rm int} \label{hamilt} \\
H_0 = & \sum_{\gamma = \uparrow,\downarrow} \int \rmd x ~\psi^{\dagger}_{\gamma} (x) \left( -\frac{\hbar^2 \partial_x^2}{2m} + U(x) - \mu \right) \psi_{\gamma}(x) \label{nonint} \\
H_{\rm int} = & \frac{1}{2} \sum_{\gamma,\beta} \int \rmd R \int \rmd r ~V(r,R) ~ \psi^{\dagger}_{\gamma}\left(R+\frac{r}{2}\right) \psi^{\dagger}_{\beta}\left(R-\frac{r}{2}\right) \nonumber \\
 & \times \psi_{\beta}\left(R-\frac{r}{2}\right) \psi_{\gamma}\left(R+\frac{r}{2}\right), \label{hint}
\end{eqnarray}
\endnumparts
where $\psi^{\dagger}_{\gamma}(x)$ creates an electron with spin
projection $\gamma$ at position $x$, and $\mu$ is the chemical
potential. We assume that the potential $U(x)$ is a smooth function of
position, and that $U(x) \ll \mu$. This allows us to introduce a
position-dependent Fermi energy $\epsilon_F (x) = \mu -
U(x)$. Similarly, the position-dependent Fermi momentum and velocity
are straightforwardly defined as $p_F(x) = \sqrt{2 m \epsilon_F(x)}$
and $v_F(x) = p_F(x)/m$.

We keep a very general form for the interaction potential between
electrons and only make the following assumptions concerning its
characteristic length scales. On the one hand, we assume for
simplicity that the interaction is short-range, the potential decaying
rapidly as a function of the distance $r$ between electrons.  On the
other hand, since we consider a non-uniform system, the interaction
depends on the position $R$ of the center of mass. The variations with
respect to $R$ are smooth and occur at a typical length scale $d$,
large compared to both the Fermi wavelength and the range of the
interaction potential. Similarly, we assume that the potential $U(x)$
varies at the same typical length scale $d$ as the interaction
strength.

\subsection{Resistivity}

We focus now on temperatures in a broad range $\hbar v_F/d \ll T \ll
\epsilon_F$. In order to compute the resistivity of the wire, we
consider a force balance on a small isolated segment of wire taken at
position $x$, whose length $\Delta x$ well exceeds the range of the
interaction while satisfying $\hbar v_F/T \ll \Delta x \ll d$. When an
external current $I$ is applied to the device, the response of the
system manifests itself as a local electric field $E(x)=\rho(x) I$,
which in turn leads to a driving force $e E(x) n(x) \Delta x$ acting
on the electrons. This driving force is compensated by a damping
force $\Delta F$ resulting from the inhomogeneous electron-electron
interaction, so that the resistivity can be written as
\begin{equation} \label{rhoofx}
\rho(x) = -\frac{\Delta F}{e n(x) I \Delta x}.
\end{equation}

\begin{figure}
\begin{center}
\includegraphics[width=12cm]{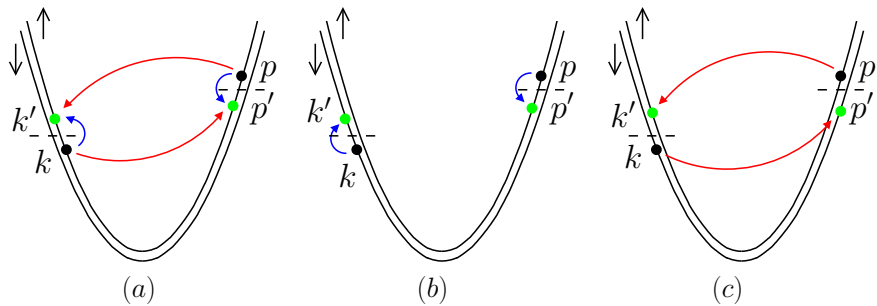}
\end{center}
\caption{The three non-momentum-conserving processes that contribute to the resistivity. These processes can be designated using the standard notations involving coupling constants \cite{solyom}: the scattering process represented in $(a)$ corresponds to $g_{2\parallel}-g_{1\parallel}$, while $(b)$ corresponds to $g_{2\perp}$, and $(c)$ to $g_{1\perp}$. Following this correspondence, in the text we use the notations $\parallel$, 2$\perp$ and 1$\perp$ to refer respectively to $(a)$, $(b)$ and $(c)$.}
\label{fig2}
\end{figure}

The damping force can be evaluated as the change in momentum per unit
time associated with two-particle scattering processes. In the regime
$k_F d \gg 1$, the processes with a large momentum difference compared
with the Fermi momentum lead to exponentially small contributions.  As a
result, in what follows we focus on processes which only weakly
violate the momentum conservation (see \fref{fig2}).

Because of the non-uniformity of the wire, strictly speaking the
momentum of the electron is not a well-defined quantity. However,
since $U(x)$ varies smoothly over a length scale $d \gg \Delta x$, it
is possible to introduce a well-defined momentum over the size of the
small segment under consideration. The expression for the momentum
thus depends on the position $x$ of the small segment, and for a state
of energy $\epsilon$ is given by $p_{\epsilon}(x) = \pm \sqrt{2m
\left[\epsilon-U(x)\right]}$. Here the $+$ sign corresponds to the
right branch, the $-$ sign to the left one.

Similarly, the eigenstates of the free Hamiltonian are no longer
given by simple plane waves but instead satisfy $\left[-\hbar^2
\partial_x^2/2m +U(x)\right] \Psi_{\epsilon}(x)=\epsilon
\Psi_{\epsilon}(x)$. Keeping in mind that the typical length scale $d$
associated with the inhomogeneities of the wire is much larger than
the Fermi wavelength, we use the semiclassical approximation, which
yields
\begin{equation}
\Psi_{\epsilon,\pm} (x)  =  \frac{1}{\sqrt{\hbar |v_{\epsilon}(x)|}} \exp \left\{ \pm \frac{\rmi}{\hbar} \int_0^x \rmd \xp \sqrt{2m \left[ \epsilon - U(\xp) \right]} \right\} \label{wavefct0}
\end{equation}
normalized according to $\int \rmd x \Psi_{\epsilon,\pm}(x)
\Psi_{\epsilon^{\prime},\pm}^{*}(x) = 2\pi \delta
(\epsilon-\epsilon^{\prime})$.  Here the velocity is defined as
$v_{\epsilon}(x)=p_{\epsilon}(x)/m$ and the index $\pm$ refers to the
right/left branches.  We ignored the backscattered
wave, since it only leads to exponentially small contributions for
$k_F d \gg 1$.

The rate of change of momentum associated with the three processes
shown in \fref{fig2} is evaluated using the Fermi golden rule, so that
the damping force acting on the electrons takes the form
\begin{eqnarray} 
\fl \Delta F =  \frac{2 \pi}{\hbar} \sum_{p,k,\pp,\kk} \left( \left|V^{\parallel}_{p k;\pp \kk}  \right|^2 +  \left|V^{2\perp}_{p k;\pp \kk}  \right|^2  +  \left|V^{1\perp}_{p k;\pp \kk}  \right|^2  \right) \delta(\epsilon_{p}+\epsilon_{k}-\epsilon_{\pp}-\epsilon_{\kk})  \nonumber \\
\times \left(\pp+\kk-p-k\right) \left[ f_p^R f_k^L (1-f_{\pp}^R) (1-f_{\kk}^L) -  f_{\pp}^R f_{\kk}^L (1-f_p^R) (1-f_k^L) \right],
\label{pdot}
\end{eqnarray}
where we introduced ${V}_{pk;\pp \kk}$ as the matrix element of the
interacting Hamiltonian \eref{hint} for scattering from the initial
state $(p,k)$ to the final state $(\pp,\kk)$ according to the
processes shown in \fref{fig2}. The superscripts $\parallel$, 2$\perp$
and 1$\perp$ refers to the standard notations for these scattering
processes \cite{solyom}. The occupation numbers $f^{R,L}$ introduced
in \eref{pdot} are given by the Fermi distribution evaluated with the
appropriate temperatures $T_{R,L}(x)$, defined in \sref{qualitative}.

One readily sees from \eref{pdot} that the damping force vanishes at
$T_R=T_L$. Using the fact that the temperature difference $T_R-T_L
\propto I$ is small in the linear response regime, we expand the
occupation numbers $f^{R,L}$ to first order in $T_R-T_L$. 
To avoid redundant derivations, let us focus on the first process
shown on \fref{fig2}$(a)$. The damping force corresponding to this
scattering process is then given by
\begin{eqnarray}
\fl \Delta F_{\parallel} =  - \frac{I}{32 \pi^2 e \epsilon_F(x)}  \int \rmd \epsilon_p \rmd \epsilon_k \rmd \epsilon_{\pp} \rmd \epsilon_{\kk}~ \left|V^{\parallel} \left(\epsilon_p, \epsilon_k;\epsilon_{\pp},\epsilon_{\kk}\right)  \right|^2 \frac{\epsilon_{\pp}-\epsilon_p + \epsilon_k - \epsilon_{\kk}}{T}\nonumber \\
\times  \left( \pp + \kk -p -k \right) \delta(\epsilon_{p}+\epsilon_{k}-\epsilon_{\pp}-\epsilon_{\kk}) 
 f_p^R f_k^L (1-f_{\pp}^R) (1-f_{\kk}^L),
\label{pdot2}
\end{eqnarray}
where we converted the summations over states into energy integrals,
and introduced the matrix element $V^{\parallel} \left(\epsilon_p,
\epsilon_k;\epsilon_{\pp},\epsilon_{\kk}\right)$ evaluated using the
set of eigenstates defined in \eref{wavefct0}.

Note that the expansion in $T_R-T_L$ leads to an expression for the
damping force which is proportional to the applied current. As a
result, in the linear response regime we can ignore any further
dependence on $I$, as this would lead to contributions that are
non-linear in the current bias. This allows us to use the Fermi energy
$\epsilon_F$ and velocity $v_F$ as they are defined in the equilibrium
state, i.e. in the reference frame where the electric current vanishes.

Focusing on states close to the Fermi energy, we can simplify the expression for the eigenstates \eref{wavefct0} of the free Hamiltonian into
\begin{equation}
\Psi_{\epsilon,\pm} (x) \simeq \Psi_{\epsilon_F,\pm} (x) \exp \left[ \pm \rmi
(\epsilon-\epsilon_F) \int_0^x \frac{\rmd \xp}{\hbar v_F(\xp)} \right] , \label{wavefct}
\end{equation}
where $\Psi_{\epsilon_F,\pm}(x)$ is obtained from \eref{wavefct0} by setting $\epsilon=\epsilon_F$. This allows us to estimate the matrix element $V^{\parallel}$ to
first order in the interaction:
\begin{eqnarray} \label{matrixelement}
\fl V^{\parallel} \left(\epsilon_p, \epsilon_k;\epsilon_{\pp},\epsilon_{\kk}\right) =  \int_x^{x+\Delta x} \rmd R ~ \frac{1}{\left[ \hbar v_F(R)\right]^2} \exp \left( \rmi \int_0^{R} \rmd \xp \frac{\epsilon_{\pp}-\epsilon_p+\epsilon_k-\epsilon_{\kk}}{\hbar v_F(\xp)} \right) \nonumber \\
\times \int_{-\Delta x}^{\Delta x} \rmd r ~ V(r, R) \left(1- \rme^{- 2\rmi k_F(R) r} \right) .
\end{eqnarray}
Here we introduced the Fermi wave vector $k_F(R)=p_F(R)/\hbar$.

Using the fact that $p$ and $\pp$ on the one hand, and, $k$ and $\kk$
on the other hand, are on the same branch, we express the momentum
difference in terms of a difference in energy by introducing the
density of states. We then define $\varepsilon= \epsilon_{\pp}-\epsilon_p+\epsilon_k-\epsilon_{\kk}$, and perform the remaining energy integrals.
Combining the resulting expression for the damping force with
\eref{rhoofx}, and substituting the matrix element
\eref{matrixelement}, we obtain the following expression for the
resistivity associated with the scattering process of \fref{fig2}$(a)$
\begin{eqnarray}
\fl \rho_{\parallel}(x) = \frac{T}{64 e^2 \epsilon_F(x) v_F(x) n(x) \Delta x} \int_x^{x+\Delta x} \rmd R_1 \frac{V_0(R_1)-V_{2k_F}(R_1)}{\pi \hbar v_F(R_1)} \nonumber \\
\times  \int_x^{x+\Delta x} \rmd R_2 \frac{V_0(R_2)-V_{2k_F}(R_2)}{\pi \hbar v_F(R_2)} \nonumber \\
\times \int \rmd \varepsilon \left[ \frac{\varepsilon/4 T}{\sinh \left( \varepsilon/4 T \right)} \right]^2~ \frac{\varepsilon^2}{\hbar v_F(R_1) \hbar v_F(R_2)} \exp \left( \rmi \varepsilon \int_{R_1}^{R_2} \frac{\rmd \xp}{\hbar v_F(\xp)} \right) . \label{pdot3}
\end{eqnarray}
The shortened notations $V_0$ and $V_{2k_F}$
correspond to the zero-momentum and $2k_F$ Fourier components of
the potential $V(r,R)$ with respect to its first variable $r$ defined as:
\begin{equation}
V_0(R) = \int \rmd r~ V(r,R) \qquad {\rm and} \qquad V_{2k_F}(R)= \int \rmd r
~V(r,R) \rme^{\rmi 2k_F(R) r}. \label{vfourier}
\end{equation}

At this stage, it is convenient to rewrite the energy integral in
\eref{pdot3} by replacing $\varepsilon^2$ with a second derivative of the
exponential term with respect to $R_1$ and $R_2$, along with the
appropriate factors of $\hbar v_F$. Performing an integration by parts
in the position variables leaves us with an expression involving single
derivatives of the dimensionless parameters $V_0(R)/\left[\pi \hbar
v_F(R)\right]$ and $V_{2k_F}(R)/\left[ \pi \hbar v_F(R)\right]$. The
remaining integral over $\varepsilon$ can be easily simplified by
noticing that it is the Fourier transform of a rapidly decaying
function which only extends over a range of energy comparable to
temperature. For temperatures $T
\gg \hbar v_F /d$, it reduces to a delta function in $R_1-R_2$ which
allows us to simplify \eref{pdot3} to
\begin{equation}
\rho_{\parallel}(x) = \frac{h}{64 e^2} \frac{T}{n(x) \epsilon_F(x)} \left[\partial_{x} \left(\frac{V_0(x)-V_{2k_F}(x)}{\pi \hbar v_F(x)} \right) \right]^2 \label{pdot4} ,
\end{equation}
where we expanded the remaining position integral to first order in
$\Delta x$.

The contributions corresponding to the remaining two scattering
processes can be computed following the same steps and are readily
obtained from \eref{pdot4} by replacing $V_0(x)-V_{2k_F}(x)$ with
$V_0(x)$ for $\rho_{2\perp}(x)$, and with $V_{2k_F}(x)$ for
$\rho_{1\perp}(x)$.  Combining the contributions from all three
processes, the final expression for the resistivity in the regime of
temperatures $T \gg \hbar v_F/d$ takes the form
\begin{eqnarray}
\fl \rho(x) = \frac{h}{64 e^2} \frac{T}{n(x) \epsilon_F(x)} \left\{ \left[ \partial_x \left( \frac{V_0(x)-V_{2k_F}(x)}{\pi \hbar v_F(x)} \right)\right]^2 \right. \nonumber \\
\left. +  \left[ \partial_x \left( \frac{V_0(x)}{\pi \hbar v_F(x)} \right)\right]^2 +  \left[ \partial_x \left( \frac{V_{2k_F}(x)}{\pi \hbar v_F(x)} \right)\right]^2 \right\} . \label{ivity}
\end{eqnarray}
This expression clearly stresses that the meaningful inhomogeneous
quantity is not just the interaction potential but rather the
dimensionless parameter that involves both the electron-electron
interaction and the Fermi velocity.  In particular, this means that a
system with a non-uniform density but homogeneous interactions between
electrons still displays a non-zero resistivity.

\section{Weakly interacting electrons in the moving frame}\label{moving}

We now introduce a different approach for evaluating the resistivity
of the system. Unlike the derivation of the previous section, this new
treatment is compatible with the bosonization formalism. Along with
providing an alternative derivation of the result \eref{ivity}, our goal
in developing this approach is to compare with the results of previous
works on the inhomogeneous Tomonaga-Luttinger liquid
\cite{maslov,safi,pono}. 

\subsection{Bosonization} \label{transfo}

Previous attempts at studying the transport properties of quantum
wires relied on an extension of the Tomonaga-Luttinger model
to inhomogeneous systems
\cite{maslov,safi,pono}. These authors assumed that the
inhomogeneities do not change the form of the Hamiltonian, and can be
accounted for by introducing position-dependent velocities and
Luttinger-liquid parameters. In the general case, however, a rigorous
derivation of the bosonized Hamiltonian for these systems is still
lacking. Here we show how such a bosonized Hamiltonian can be derived
explicitly in the case of a non-uniform system of weakly interacting
electrons.

The standard bosonization formula for weakly interacting fermions
involves the Fermi momentum as well as a momentum cutoff (see
e.g. \cite{LL-th}), and as such cannot be straightforwardly extended
to non-uniform systems where both these quantities can develop a
position dependence. The key idea then is to map the inhomogeneous
system of electrons onto a set of fictitious fermions described by a
Hamiltonian whose non-interacting part is translationally
invariant. From there, a standard bosonization procedure holds and the
resulting Hamiltonian expressed in terms of the new variables is very
reminiscent of the conjectured inhomogeneous Luttinger liquid
Hamiltonian, in the limit of weak interactions (see \sref{discussion}).

Our starting point is similar to the one we considered in
\sref{stationary}, namely a system of interacting electrons with a
non-uniform density described by the Hamiltonian \pref{hf}. As we
noticed in the previous section, the non-uniform potential $U(x)$
appearing in \eref{nonint} breaks the translational invariance of the
system, already when no electron-electron interaction is present. As a
result, the eigenstates of the free Hamiltonian are no longer plane
waves but for energies close to the Fermi level, they can be
approximated by \eref{wavefct}.

Up to a prefactor which depends on position but not on $\epsilon$,
these low-energy eigenstates look like plane waves, putting forth a
more natural set of variables: the energy difference
$\epsilon-\epsilon_F$ and $X(x) = \int_0^x \rmd \xp / \hbar
v_F(\xp)$. An expansion of the electron field operator
$\psi_{\gamma,\pm}(x)$ over these plane waves calls for the
introduction of a fictitious fermion field operator
$\eta_{\gamma,\pm}(X)$ defined as
\begin{equation}
\psi_{\gamma,\pm} (x) = \Psi_{\epsilon_F,\pm} (x)~ \eta_{\gamma,\pm} \left(X(x)\right) , \label{defeta}
\end{equation}
where $\Psi_{\epsilon_F,\pm}(x)$ was introduced in \eref{wavefct}.
Note that the anti-commutation relations satisfied by
$\psi_{\gamma,\pm}(x)$ transfer to $\eta_{\gamma,\pm}(X)$ ensuring that
$\left\{
\eta_{\gamma,\sigma}(X),\eta_{\beta,\sigma^{\prime}}^{\dagger}(Y)\right\}=
\delta_{\sigma \sigma^{\prime}} \delta_{\gamma \beta} \delta(X-Y)$.

Let us now derive the Hamiltonian describing the physics of these
fictitious fermions. This is accomplished by substituting
\eref{defeta} into the Hamiltonian \pref{hf}. By construction, the
free Hamiltonian is translationally invariant in the new variable $X$. At
low energy, the interacting part of the Hamiltonian can be decomposed
in three sectors corresponding to the conventional $g_1$, $g_2$ and
$g_4$ processes \cite{solyom}. The main difference here is that the
associated coupling constants are now position-dependent. They can be
obtained from the Fourier components of the electron-electron
interaction potential. 

As an example, consider the so-called $g_{2 \parallel}$
process. Following \cite{LL-th}, the coupling constant for this
process is given by the zero-momentum Fourier component of the
interaction potential, which in the case of our inhomogeneous system
corresponds to $V_0(x)$, introduced in \eref{vfourier}.  Replacing
$\psi$ with $\eta$ according to \eref{defeta}, and introducing the
density operator $\nu_{\gamma,\pm}(X)=\eta_{\gamma,\pm}^{\dagger} (X)
\eta_{\gamma,\pm} (X)$, the $g_{2 \parallel}$ process retains the same
form
\begin{equation}
\int \rmd x ~ g_{2 \parallel} (x) \rho_{\gamma,\sigma}(x) \rho_{\gamma,-\sigma}(x) \longrightarrow \pi \int \rmd X~ y_{2 \parallel}(X) \nu_{\gamma,\sigma}(X) \nu_{\gamma,-\sigma}(X) 
\end{equation}
only with a dimensionless coupling constant given by $y_{2
\parallel}\left(X(x)\right) = g_{2 \parallel}(x)/\pi\hbar v_F(x)$. A
similar treatment can be applied to the remaining sectors of the
interaction.

The resulting Hamiltonian expressed in terms of the fictitious field
$\eta$ can now be bosonized following the standard procedure:
\numparts
\plabel{formula}
\begin{eqnarray}
\eta_{\uparrow,\pm} (X) = \frac{U_{\gamma,\pm}}{\sqrt{2\pi \alpha}} \exp \left\{ \frac{-\rmi}{\sqrt{2}} \left[ \pm \phi_{\rho}(X)-\theta_{\rho}(X) \pm \phi_{\sigma}(X)- \theta_{\sigma}(X)\right] \right\} \\
\eta_{\downarrow,\pm} (X) = \frac{U_{\gamma,\pm}}{\sqrt{2\pi \alpha}} \exp \left\{ \frac{-\rmi}{\sqrt{2}} \left[ \pm \phi_{\rho}(X)-\theta_{\rho}(X) \mp  \phi_{\sigma}(X) + \theta_{\sigma}(X)\right] \right\}
\end{eqnarray}
\endnumparts
where we introduced the fields $\phi_{\nu}$ and $\theta_{\nu}$ (with
$\nu=\rho,\sigma$) satisfying bosonic commutation relations
$\left[\phi_{\nu}(X),\partial_Y \theta_{\nu}(Y)\right] = \rmi \pi
\delta(X-Y)$. Here $U_{\gamma,\pm}$ are the standard Klein factors
\cite{LL-th} and $\alpha^{-1}$ is an energy
cutoff\footnote{Considering that the fictitious fermions were
introduced in the vicinity of the electron Fermi surface, one should
assume $\alpha^{-1} \ll \epsilon_F$.} introduced to regularize the
theory in the ultra-violet sector.

In terms of the bosonic variables, the Hamiltonian of the system can
be written as a sum of two terms describing the excitations of charge
and spin degrees of freedom respectively, and takes the form
\numparts
\plabel{hb}
\begin{eqnarray}
H & = & H_{\rho} + H_{\sigma} \label{hboso} \\
H_{\rho} & = & \frac{1}{2\pi} \int \rmd X ~ \left[ \left( \partial_X \theta_{\rho}\right)^2 + \left( 1 + y_{\rho}(X) \right) \left( \partial_X \phi_{\rho} \right)^2 \right] \label{hrho} \\
H_{\sigma} & =& \frac{1}{2\pi} \int \rmd X ~ \left[ \left( \partial_X \theta_{\sigma}\right)^2 + \left( 1 - y_{\sigma} (X) \right) \left( \partial_X \phi_{\sigma} \right)^2 \right] \nonumber \\
& & + \frac{2}{\left( 2\pi \alpha \right)^2} \int \rmd X ~y_{\sigma}(X) \cos \left( 2\sqrt{2} \phi_{\sigma}\right) . \label{hsigma}
\end{eqnarray}
\endnumparts
The dimensionless parameters $y_{\rho}$ and $y_{\sigma}$ are
conventional notations for combinations of $y_1$, $y_2$ and $y_4$
\cite{LL-th} given by
\begin{equation}
y_{\rho} \left(X(x)\right) = \frac{V_{2k_F}(x)-2 V_0(x)}{\pi \hbar v_F (x)} \qquad y_{\sigma} \left(X(x)\right) = \frac{V_{2k_F}(x)}{\pi \hbar v_F(x)} , \label{yrhosigma}
\end{equation}
where $V_{0}$ and $V_{2k_F}$ are the Fourier components of the
interaction potential as defined in \eref{vfourier}.

Note that this form of bosonization clearly highlights that the
important variables are the dimensionless parameters $y_{\rho,\sigma}$
rather than the various interaction constants $g_{1,2,4}$. The
calculation of the resistivity carried out in \sref{stationary} led to
the same observation, see \eref{ivity}.

\subsection{Resistance and dissipation} \label{dissipate}

A way to determine the resistivity of the system is to relate it to
the mechanism of dissipation of energy into the wire when an external
current bias is applied. This relation was explored in \cite{matveev}
in the context of a quantum wire in the Wigner crystal regime, and the
method we outline here is similar.

In the presence of an applied current $I=I_0 \cos \omega t$, the
electrons start moving in the wire. More specifically, in the dc limit
$\omega \to 0$, one can assume that the current is uniform throughout
the wire and all electrons move in phase. As a result, the position of
the electrons depends on time and is related to the injected charge
$q(t)=I_0 \omega^{-1} \sin
\omega t$ defined as $I(t)= \dot{q}(t)$. This time dependence of the
positions of the electrons can be accounted for by replacing $x \to
x+q(t)/e n(x)$ in the position-dependent parameters of the
Hamiltonian.  While this has no effect in practice when the
translational invariance holds, for an inhomogeneous system it leads
to a time-dependent perturbation to the Hamiltonian. Alternatively,
this amounts to describing the system in the reference frame moving
with the electron fluid. In this case, the electrons experience the
effect of an inhomogeneous potential moving as a function of time.

In terms of the fictitious set of fermions $\eta_{\gamma,\pm} (X)$, one
needs to substitute $X$ in the dimensionless interaction parameters
$y_{\rho,\sigma}$ by the time-dependent position $X+q(t)/e n(X)$
where the density in these variables is given by $n(X(x))= n(x)
\hbar v_F(x)$. In the linear response regime, an expansion to first order in
$q(t)$ leads to the following form of the Hamiltonian:
\begin{equation} \label{newhint}
H \longrightarrow H + \int \rmd X \frac{q(t)}{e n(X)} {\cal H}^{\prime}(X) ,
\end{equation}
where we introduced the notation ${\cal H}^{\prime}(X)={\cal
H}_{\rho}^{\prime}(X)+{\cal H}_{\sigma}^{\prime}(X)$ for the following
quantities:
\numparts
\begin{eqnarray}
{\cal H}_{\rho}^{\prime} (X) = \frac{1}{2 \pi} \left[\partial_X y_{\rho}(X)\right] \left( \partial_X  \phi_{\rho} \right)^2 \label{pert_charge}\\
{\cal H}_{\sigma}^{\prime} (X) = - \frac{1}{2 \pi} \left[\partial_X y_{\sigma}(X)\right] \left( \partial_X  \phi_{\sigma} \right)^2 + \frac{2}{(2\pi \alpha)^2} \left[\partial_X y_{\sigma} (X)\right] \cos (2 \sqrt{2} \phi_{\sigma}) . \label{pert_spin}
\end{eqnarray}
\endnumparts

The time-dependent perturbation in \eref{newhint} acts as an external
driving force, resulting in the creation of spin and charge
excitations. These excitations are responsible for dissipating the
energy from the external force into the wire. Using the Fermi golden
rule, it is possible to estimate the rates of these absorption and
emission processes, and therefore, the energy $W$ dissipated in unit
time into the system. In the linear response regime, where the
amplitude $I_0$ of the current oscillations is weak, the energy
dissipated in unit time is quadratic in $I_0$ and is given by
\begin{eqnarray} 
\fl W = \hbar \omega \frac{2\pi}{\hbar} \left( \frac{I_0}{2 e \omega} \right)^2 \int_{-\infty}^{+\infty} \frac{\rmd t}{2 \pi \hbar} \left( \rme^{\rmi \omega t} - \rme^{-\rmi \omega t}\right) \nonumber \\
\times \int \frac{\rmd X}{n(X)} \int \frac{\rmd Y}{n(Y)} \langle {\cal H}^{\prime} (X,t) {\cal H}^{\prime} (Y,0)\rangle ,
\label{dissip}
\end{eqnarray}
where $\langle \ldots \rangle$ stands for thermodynamic averaging.

The resistance of the system is then derived by comparing the
dissipated energy obtained in \eref{dissip} with the Joule heat law
$W=I_0^2 R /2$. Since the charge part \eref{pert_charge} and the spin
part \eref{pert_spin} of the time-dependent perturbation commute, one
expects the resistance $R$ of the wire to be expressed as the sum of a
spin and a charge contribution $R=R_{\rho}+R_{\sigma}$, which can be
evaluated separately. This can be understood as the consequence of
having two independent channels for dissipating energy throughout the
wire, corresponding to spin and charge excitations
\cite{matveev}. After some manipulations, these two contributions to
the resistance can be expressed in the dc limit as
\begin{equation} \label{resistance}
R_{\nu} = - \frac{1}{\hbar e^2}~ \int \frac{\rmd X}{n(X)} \int
\frac{\rmd Y}{n(Y)} \lim_{\omega \to 0} \frac{{\rm Im}\left[{\cal
W}_{{\rm ret},\nu} (X,Y;\omega)\right]}{\omega} .
\end{equation}
Here we introduced the retarded correlator ${\cal W}_{{\rm ret},\nu}$ as the Fourier transform in time of ${\cal W}_{{\rm ret},\nu} (X,Y;t) =
-\rmi \theta(t) \langle \left[{\cal H}_{\nu}^{\prime}(X,t),{\cal H}_{\nu}^{\prime}(Y,0)\right]
\rangle$, where $\nu=\rho,\sigma$.

\subsection{Charge contribution to the resistivity} 

In order to derive the contribution to the resistivity from charge
degrees of freedom, we substitute in \eref{resistance} the expression
for ${\cal H}_{\rho}^{\prime}$ introduced in \eref{pert_charge}. The
retarded correlator ${\cal W}_{{\rm ret},\rho}$ resulting from this
substitution is quartic in the bosonic field $\phi_{\rho}$. It is thus
more convenient to express it in terms of the corresponding
time-ordered correlation function, via an analytic continuation in
frequency space. This allows for the use of Wick's theorem, ultimately
leading to the following expression of the retarded correlator:
\begin{eqnarray}
\fl {\cal W}_{{\rm ret},\rho} \left( X,Y;\omega\right) = -  \frac{\hbar}{2\pi^2} \left[ \partial_X y_{\rho}(X)\right] \left[ \partial_Y y_{\rho}(Y)\right] \nonumber \\
\times \left[ \int_0^{\beta} \rmd \tau \rme^{\rmi \nu_n \tau} \left( \frac{\partial^2}{\partial X \partial Y} \langle T \phi_{\rho}(X,\tau) \phi_{\rho}(Y,0) \rangle \right)^2 \right]_{i\nu_n \to \hbar \omega+i\delta} , \label{corr}
\end{eqnarray}
where $\langle T \ldots \rangle$ corresponds to the time-ordered correlation function.

Since \eref{corr} is explicitly quadratic in the interaction, the
dominant contribution to the retarded correlator ${\cal W}_{{\rm
ret},\rho}$ can be derived by using the free propagator of the bosonic
field $\phi_{\rho}$. The latter is readily obtained from the
non-interacting Hamiltonian, and is given by
\begin{equation}
\langle T \phi_{\rho} (X,\tau) \phi_{\rho}(Y,0) \rangle = T \sum_n \int \frac{\rmd K}{2 \pi} \frac{\pi}{\nu_n^2 + K^2} \rme^{\rmi K (X-Y)} \rme^{-\rmi \nu_n \tau} .  \label{propagator}
\end{equation}
Combining \eref{propagator} with \eref{corr}, performing the integral
over imaginary time $\tau$, and substituting the analytically
continued result into \eref{resistance}, the charge contribution to
the resistance in the dc limit reads:
\begin{eqnarray}
\fl R_{\rho} = - \frac{\pi \hbar T^2}{8 e^2} \int \rmd X \frac{\partial_X y_{\rho}(X)}{n(X)} \int \rmd Y \frac{\partial_Y y_{\rho} (Y)}{n(Y)} \nonumber \\
\times \frac{1}{\sinh^2 \left[ 2\pi T (X-Y) \right]} \left\{ 1 - \frac{2\pi T (X-Y)}{\tanh \left[ 2\pi T (X-Y)\right]}\right\} \label{srkernel} ,
\end{eqnarray}
where we restricted ourselves to contributions up to second order in
the interaction.

One recognizes in \eref{srkernel} a rapidly decaying integral
kernel for $|X-Y| \gg 1/T$. In terms of real space quantities,
this corresponds to distances of order $\hbar v_F / T$. It follows
that at temperatures $T \gg \hbar v_F/d$ the double integral in $X$
and $Y$ is dominated by short-range contributions. This allows us to
reduce the expression \eref{srkernel} for the resistance to a single
integral over $X$. Changing variables back from $X$ to $x$, the
integrand of the resulting expression for the resistance $R_{\rho}$
can be identified with the charge contribution to the resistivity at
position $x$ in space and is given by
\begin{equation}
\rho_{\rho} (x) = \frac{h}{128 e^2} \frac{T}{\epsilon_F(x) n(x)} \left[ \partial_x y_{\rho}(x) \right]^2 ,\label{rhorho}
\end{equation}
where we focused on temperatures in the range $\hbar v_F /d \ll T \ll
\epsilon_F$.

\subsection{Spin contribution to the resistivity}

The method used above to derive the charge contribution to the
resistivity can be readily extended to evaluate that of spin degrees
of freedom. Substituting the expression for ${\cal
H}_{\sigma}^{\prime}$ into \eref{resistance}, one notices that the
spin contribution splits off into two parts: one coming from the
quadratic term in $\phi_{\sigma}$, the other from the cosine term.

\subsubsection{Quadratic term}

The expressions for the quadratic parts of ${\cal H}_{\rho}^{\prime}$
and ${\cal H}_{\sigma}^{\prime}$ are identical up to a sign, upon
replacing the charge parameter $y_{\rho}$ and field $\phi_{\rho}$ by
their spin counterparts. As a result, in order to derive the
contribution to the transport properties from the quadratic term in
$\phi_{\sigma}$, it is sufficient to repeat the steps leading to \eref{rhorho} but replace $y_{\rho}$ by $y_{\sigma}$ so that
\begin{equation} \label{rhosigmaquad}
\rho_{\sigma,{\rm quad}} (x) = \frac{h}{128 e^2} \frac{T}{\epsilon_F(x) n(x)} \left[ \partial_x y_{\sigma}(x) \right]^2 .
\end{equation}
Here we again restricted ourselves to temperatures in the range $\hbar
v_F/d \ll T \ll \epsilon_F$.

\subsubsection{Cosine term}

The contribution to the resistivity coming from the cosine term of the
spin Hamiltonian can be inferred from \eref{rhosigmaquad} based on the
following symmetry argument. In terms of the bosonized Hamiltonian,
the interaction-dependent term appearing in the quadratic part of
$H_{\sigma}$ accounts for the coupling between $z$ components of the
electron spins \cite{LL-th}. This term ultimately leads to the
contribution $\rho_{\sigma,{\rm quad}}$ obtained in
\eref{rhosigmaquad}. On the other hand, the cosine term in
\eref{hsigma} corresponds to the coupling of the remaining $x$ and $y$
components \cite{LL-th}. Because of the SU$(2)$ symmetry, the
contributions from all three components of the interaction between
electron spins are the same. As a result, we expect the cosine term to
contribute twice as much to the resistivity as the quadratic part of
the spin Hamiltonian.

This can be verified explicitly by substituting the cosine term from
\eref{pert_charge} into the expression for the retarded correlator
${\cal W}_{{\rm ret},\sigma}$. After performing the analytic
continuation and taking the dc limit $\omega \to 0$, the cosine-cosine
correlation function takes the form
\begin{eqnarray}
\fl \lim_{\omega \to 0} \frac{1}{\omega} {\rm Im} \left\{ \left[ \int_0^{\beta} \rmd \tau \rme^{\rmi \nu_n \tau} \langle T \cos \left( 2\sqrt{2} \phi_{\sigma}(X,\tau)\right) \cos \left(2\sqrt{2} \phi_{\sigma}(Y,0) \right)\rangle \right]_{\rmi \nu_n \to \hbar \omega+\rmi \delta} \right\}  \nonumber \\
=- \frac{\pi^3 \hbar T^2 \alpha^4}{\sinh^2 \left[ 2\pi T (X-Y) \right]} \left\{ 1 - \frac{2\pi T (X-Y)}{\tanh \left[2\pi T (X-Y)\right]}\right\} . \label{coscos}
\end{eqnarray}
Here it is sufficient to perform the thermodynamic averaging using the
free Hamiltonian since this term enters ${\cal W}_{{\rm ret},\sigma}$
with a prefactor quadratic in the interaction parameter $y_{\sigma}$.

One recognizes in \eref{coscos} the same short-range kernel we encountered in
\eref{srkernel}. It follows that, at temperatures $T \gg \hbar
v_F/d$, one can simplify the expression for the resistance into a
single integral over $X$. Changing back variables from $X$ to $x$, and
identifying the integrand in $x$ with the resistivity, we obtain the
contribution from the cosine term $\rho_{\sigma,{\rm cos}}(x) = 2
\rho_{\sigma,{\rm quad}}(x)$, as we argued from the SU$(2)$ symmetry.
The total contribution from spin degrees of freedom thus amounts to
three times the result \eref{rhosigmaquad}.

Combining the charge and spin contributions, the final expression for the resistivity of the wire at temperatures $T \gg \hbar v_F/d$ is given by
\begin{equation}
\rho (x) =  \frac{h}{128 e^2} \frac{T}{\epsilon_F(x) n(x)} \left\{ \left[ \partial_x y_{\rho}(x)\right]^2 + 3 \left[\partial_x y_{\sigma}(x)\right]^2  \right\} . \label{finalrho}
\end{equation}
Using \eref{yrhosigma} to replace the dimensionless parameters
$y_{\rho}(x)$ and $y_{\sigma}(x)$ with their expression in terms of
the electron-electron interaction potential, the latter result becomes
identical to \eref{ivity}.

\section{Discussion}\label{discussion}

\subsection{Inhomogeneous Luttinger liquid}

It is interesting to compare the bosonized Hamiltonian we derived in
\pref{hb} to that of the inhomogeneous Tomonaga-Luttinger model
conjectured in \cite{maslov,safi,pono}. To do so, we change variables
back from $X$ to $x$, so that the Hamiltonian \pref{hb} takes the
form
\numparts
\begin{eqnarray}
H & = & H_{\rho} + H_{\sigma} \\
H_{\rho} & = & \int \rmd x ~\frac{\hbar v_F(x)}{2\pi} \left[ \left( \partial_x \vartheta_{\rho}\right)^2 + \left( 1 + y_{\rho}(x) \right) \left( \partial_x \varphi_{\rho} \right)^2 \right]  \\
H_{\sigma} & =& \int \rmd x ~\frac{\hbar v_F(x)}{2\pi} \left[ \left( \partial_x \vartheta_{\sigma}\right)^2 + \left( 1 - y_{\sigma} (x) \right) \left( \partial_x \varphi_{\sigma} \right)^2 \right] \nonumber \\
& & + \int \rmd x ~ \frac{2 g_{\sigma}(x)}{\left[2 \pi \alpha \hbar v_F(x)\right]^2} \cos \left( 2\sqrt{2} \varphi_{\sigma}\right) ,
\end{eqnarray}
\endnumparts
where we denoted $\varphi_{\nu}(x)=\phi_{\nu}(X(x))$ and
$\vartheta_{\nu}(x)=\theta_{\nu}(X(x))$ (with $\nu=\rho,\sigma$).

The charge Hamiltonian $H_{\rho}$ and the quadratic part of
$H_{\sigma}$ are identical to the inhomogeneous Tomonaga-Luttinger
Hamiltonian \cite{maslov,safi,pono}, taken in the limit of weak
interactions. The important difference comes from the cosine term of
the spin Hamiltonian. This term was absent from previous works which
either discarded it arguing that the coupling constant $g_{\sigma}$
renormalizes towards zero at low energy scales \cite{ITLL}, or simply
focused on a system of spinless fermions \cite{maslov,safi,pono}. To
recover a standard form for the cosine term, one needs to introduce a
position-dependent momentum cutoff $\left[ \alpha(x) \right]^{-1}$
defined as $\alpha(x) = \alpha \hbar v_F(x)$, where $\alpha^{-1}$ is
the energy cutoff introduced in \pref{formula}. Keeping in mind that
momentum is no longer a conserved quantity in our model, this position
dependence of the momentum cutoff was to be expected. Interestingly
though, the natural guess relying on the common interpretation of
$\alpha(x)$ as a small distance cutoff, would have led to a different
answer. Indeed, assuming that $\alpha(x)$ represents the shortest
inter-particle distance, one would expect it to be inversely
proportional to the electron density, i.e. $\alpha(x) \propto
1/v_F(x)$.

Although the quadratic part of our bosonized Hamiltonian is similar to
the model considered \cite{maslov,safi,pono}, we do not reach the same
final answer. This is because our treatment amounts to considering
perturbations to the Luttinger-liquid Hamiltonian which were not taken
into account in previous studies. In \sref{moving}, we treated $I(t)$
as an external parameter. In the framework of the Luttinger liquid
model, it can also be interpreted as an excitation of the charge
mode. Using the bosonization expression for the electric current
$I=e(\sqrt{2}/\pi) \dot{\phi_{\rho}}$, one readily sees that $q(t)$
can thus appear as a dynamical variable, directly proportional to the
charge field $\phi_{\rho}$. As a result, the linear in $q(t)$
perturbation to the Hamiltonian in \eref{newhint} corresponds, in the
conventional Luttinger-liquid theory, to cubic terms in the bosonic
fields of the form $\phi_{\rho} (\partial_X \phi_{\nu})^2$
($\nu=\rho,\sigma$). These terms are irrelevant perturbations to the
Luttinger-liquid Hamiltonian, and as such are usually discarded.
However, it was proven that within the quadratic Luttinger-liquid
Hamiltonian, non-uniform electron-electron interactions do not
contribute to the resistance \cite{maslov,safi,pono}. It thus makes
sense to take these irrelevant perturbations into account. Our
approach showed that they affect the transport properties in a
non-trivial way and lead to a finite resistivity.

\subsection{Connection with experiments}

Our results are relevant to experiments performed on long quantum
wires. However, we focused on the case of weakly interacting electrons
which is unlikely to be realized in experimental situations.  Let us
discuss to what extent our conclusions are modified when this
restriction on the interaction strength is relaxed.

Though our results are not readily applicable in the case of strong
electron-electron interactions, the method developed in \sref{moving}
which relies on bosonization suggests that the temperature and density
dependences of the resistivity should not be affected by the strength
of the interactions. Experimental measurements of these dependences
may thus be compared with our results.

Furthermore, given the Hamiltonian of the system in the
strongly interacting regime, one could repeat the treatment of
\sref{moving} in order to derive the resistivity. Unfortunately, a
rigorous derivation of the bosonized Hamiltonian in the case of a
strongly interacting inhomogeneous system is yet to be found.

\subsection{Equilibration}

In our derivation, we assumed that the electronic subsystem is in
equilibrium in the moving frame. For this to be satisfied, we need the
wires to be longer than the typical length scale $l_{\rm eq}$
associated with the processes of equilibration taking place inside the
wire. If the size of the wire becomes too short with respect to the
equilibration length $l_{\rm eq}$, we expect our results to be
modified by an additional small prefactor of the order of the ratio of
these two length scales. This might lead to a non-trivial temperature
dependence of the resistance, depending on the leading equilibration
mechanism involved.

Little is known about equilibration mechanisms in one-dimensional
interacting systems. In the case of weakly interacting electrons,
recent work \cite{3electron} suggests that scattering processes
involving three electrons may be the leading source of equilibration
in the system. Because of consideration of energy and momentum
conservation, these three-particle collisions should involve states
near the bottom of the band, resulting in a strong suppression at low
temperatures. In the experimentally relevant case of low electron
density and strong interactions, this analysis no longer holds and a
detailed treatment remains elusive. It is natural to expect that the
equilibration in the wire would become easier as the interactions grow
stronger.

\section{Summary}\label{conclusion}

In this paper we studied the effect of inhomogeneous electron-electron
interactions on the transport properties of a quantum wire. We
considered a very general form of the interaction potential, and
allowed for a non-uniform density of electrons along the wire. We
argued that the inhomogeneities allow for non-momentum-conserving
scattering processes which give rise to a finite resistivity of the
wire. We showed that in the regime of weakly interacting electrons,
such scattering processes contribute to the resistivity as a linear in
$T$ term\footnote{In the low temperature regime $T \ll \hbar
v_F/d$, our preliminary results suggest a much weaker $T^4$ contribution
to the resistance.}, over a broad range of temperatures $T$ below the
Fermi energy. We also reformulated our results within the framework of
the inhomogeneous Tomonaga-Luttinger model, and analyzed the
differences with previous works relying on this formalism.

\ack
We are grateful to A. V. Andreev, T. Giamarchi and L. I. Glazman for
helpful discussions.  This work was supported by the U.S. Department
of Energy, Office of Science, under Contract No. DE-AC02-06CH11357.

\end{document}